\documentclass[twocolumn,showpacs,preprintnumbers,showkeys,superscriptaddress]{revtex4}
\usepackage{graphicx}
\usepackage{dcolumn}
\usepackage{bm}

\def\eqref#1{Eq.~(\ref{eq:#1})}

\begin{document}

\title{Strong Linear Correlation Between Eigenvalues and Diagonal Matrix Elements}
\author{J. J. Shen}
\affiliation{Department of Physics,  Shanghai Jiao Tong University,
Shanghai 200240, China}
\author{A. Arima}
\affiliation{Department of Physics,  Shanghai Jiao Tong University,
Shanghai 200240, China}
\affiliation{Science Museum, Japan Science
Foundation, 2-1 Kitanomaru-koen, Chiyoda-ku, Tokyo 102-0091, Japan}
\author{Y. M. Zhao}    \email{ymzhao@sjtu.edu.cn}
\affiliation{Department of Physics,  Shanghai Jiao Tong University,
Shanghai 200240, China}  \affiliation{Center of Theoretical Nuclear
Physics, National Laboratory of Heavy Ion Accelerator, Lanzhou
730000, China} \affiliation{CCAST, World Laboratory, P.O. Box 8730,
Beijing 100080, China}
\author{N. Yoshinaga}    \email{yoshinaga@phy.saitama-u.ac.jp}
\affiliation{Department of Physics, Saitama University, Saitama
338-8570, Japan}

\date{\today}

\begin{abstract}
We investigate eigenvalues of many-body systems interacting by
two-body forces as well as those of random matrices. We find a
strong linear correlation between eigenvalues and diagonal matrix
elements if both of them are sorted from the smaller values to larger
ones. By using this linear correlation we are able to predict
reasonably all eigenvalues of given shell model Hamiltonian without
complicated iterations.
\end{abstract}


\vspace{0.4in}

\maketitle

\newpage

Diagonalization of matrices is a very common practice in many
fields. There have been a number of algorithms for diagonalization,
such as the Jacobi method,  the Householder method, the Lanczos
method, etc. \cite{Numerical}.  Full diagonalizations
of obtaining all the eigenvalues
become
difficult when the dimension of   matrices is larger than $10^{5}$.
One can obtain   a very few lowest eigenvalues of sparse matrices
when dimension of  matrices is $10^{8-10}$ by super computers. There
have been also many efforts in obtaining eigenvalues by using the
energy centroid and spectral moments, and typical works along this
line can be found in, e.g.,  Refs.
\cite{Ratcliff,Zuker,Vary,Papenbrock,Yoshinaga}. Unfortunately, the
eigenvalues such obtained by using centroids and moments are valid
at the statistical level (i.e., {\it after the ensemble average})
and not good enough for individual diagonalizations.

The purpose of this Letter is to report a remarkable correlation
between eigenvalues and diagonal matrix elements, if both are sorted
from smaller values to larger ones. One can predict approximately
all eigenvalues of given matrices by using this correlation.
Although eigenvalues such predicted are not exact, they are good
approximations of exact results (within a precision of $1-4\%$ in
examples that we have studied).

Let us begin with Hamiltonian matrices which conserve the angular momentum.  The
two-body interaction parameters are taken to be Gaussian random
numbers  (i.e., two-body random ensemble \cite{French} which is
called the TBRE). In such cases both the diagonal matrix elements
of a many-body system and their eigenvalues are known to exhibit Gaussian
distributions approximately. One easily conjectures that there might exist a
linear correlation between these two sets of quantities. Let us
denote exact eigenvalues obtained by diagonalization by using
$E_i^{\rm exact}$, dimension of matrix by   $D$ (number of spin $I$
states is denoted by $D_I$), predicted eigenvalues by  $E_i^{\rm
exact}$, diagonal matrix elements by   $H_{ii}$, and average energy
by $\overline{H}$. We define the linear correlation coefficient $r=
\frac{ \sum_i (H_{ii} - \overline{H}) (E_i^{\rm exact}-
\overline{H})}{ \sqrt{ \sum_i (E_i^{\rm exact}-\overline{H})^2
\sum_i (H_{ii} - \overline{H})^2 } }$. The absolute value of $r$, $|r|$, is less than
or equals to 1. If there exists a strong linear correlation between
$E_i^{\rm exact}$ and $H_{ii}$ ($i=1, 2, \cdots n$), $|r|
\rightarrow 1$. We investigate below whether or not the coefficient
$r$ is very close to 1 by a number of different configurations. We
take six successive sets of random interactions for each
system in demonstrating the correlation between eigenvalues
and diagonal matrix elements.

In Fig. 1(a-f), we present exact eigenvalues versus diagonal matrix
elements of $I=20$ states (dimension $D_I=29$) for four identical
fermions in a single-$j$ ($j=\frac{31}{2}$) shell. Both eigenvalues
and diagonal matrix elements are sorted from the smallest to the
largest. The linear correlation coefficients $r$ are more than 0.95 in
all these panels. Each panel corresponds to one set of random
interaction parameters.

In Fig. 2(a-f), similar results are presented for $sd$ boson systems
 \cite{IBM}. Here $I=6$ and boson number $n=12$ (dimension $D_I=37$).
In Fig. 3(a-f) we present results for the $I=17$ ($D_I=508$) states
of three valence protons and three valence neutrons in the
$(2s_{1/2},1d_{3/2},0i_{11/2})$ shell; in Fig. 4 we present results
of  $I=4$ ($D_I=3017$) states of the same configuration. The
Hamiltonian that we use in Figs. 3-4 conserves parity and isospin as
well as angular momentum, i.e., here we use the conventional shell
model Hamiltonian in nuclear structure physics.  The two-body
interactions   are taken to be random values following the Gaussian
distribution. Correlations in all these figures are seen to be
striking, although there are small deviations when eigenvalues
are around the smallest or the largest.

Now we discuss how one obtains the predicted linear correlation by a
simple algorithm.  Let us assume that
\begin{equation}
E_{i}=A H_{ii}+B, \label{eq1}
\end{equation}
where $E_{ii}$ is the eigenvalues of the Hamiltonian. By using
\begin{eqnarray}
&&  \sum_{i=1}^D E_i=\sum_{i=1}^D H_{ii}, \nonumber \\
&&  \sum_{i=1}^D (E_i)^2=\sum_{i,j=1}^D H_{ij}^2, \label{eq2}
\end{eqnarray}
we obtain
\begin{eqnarray}
&& A=\sqrt{\frac{D\sum_{i=1}^D \sum_{j=1}^D H_{ij}^2-(\sum_{i=1}^D
H_{ii})^2}{D\sum_{i=1}^D
H_{ii}^2-(\sum_{i=1}^D H_{ii})^2}} \nonumber \\
&& ~~~~~~~~  =  \sqrt{ \frac{ \overline{H^2}-\overline{H}^2}
{\sum_i H^2_{ii}/D-\overline{H}^2} } , \nonumber \\
&& B=  (1-A) \overline{H} .
\end{eqnarray}
All straight lines in this paper are plotted by using the above
relation.

\begin{center}
\includegraphics[width = 3in]{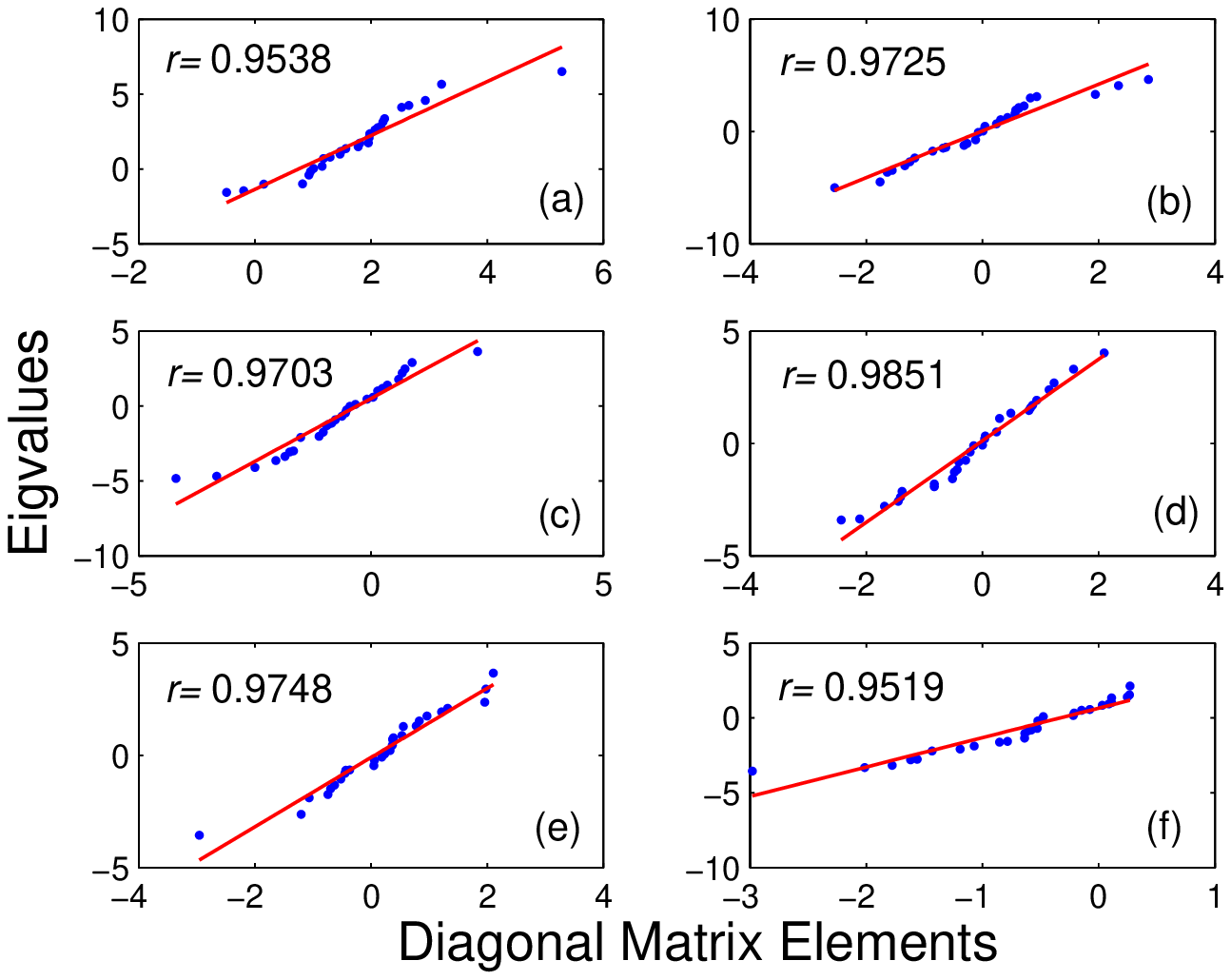}
\end{center}

FIG. 1. ~(Color online)~ Linear correlation between exact
eigenvalues and diagonal matrix elements. These examples correspond
to four fermions in a single-$j$ shell with $j=31/2$. Here $I=20$,
$D_I$=29. $r$ is the linear correlation coefficient. Straight lines
are plotted by using Eq. (1), with $A$ and $B$ values evaluated by Eq.
(3).

\vspace{0.1in}

\begin{center}
\includegraphics[width = 3in]{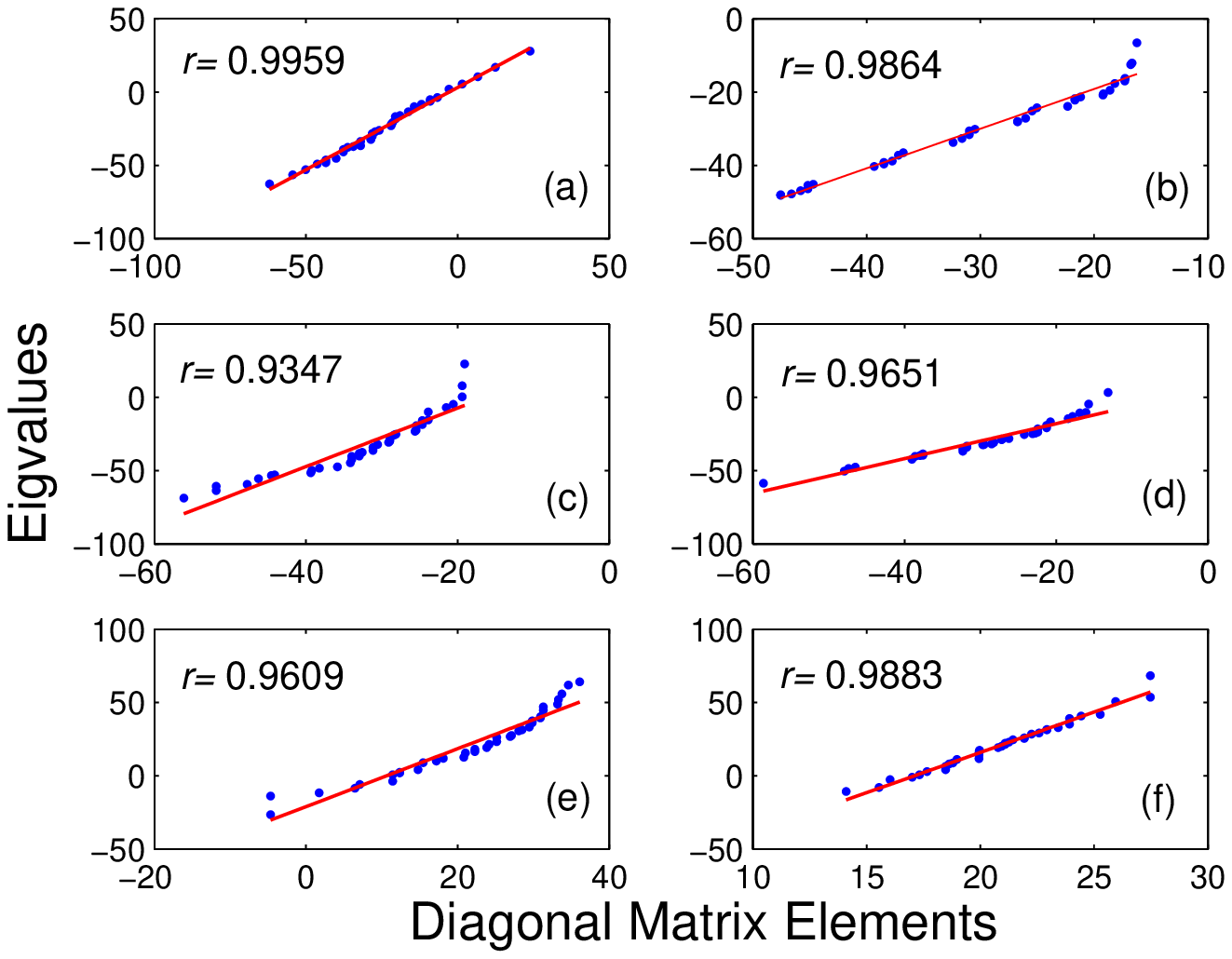}
\end{center}

FIG. 2. ~(Color online)~ The same as Fig. 1 except that these
examples correspond to  $sd$-boson systems. Here boson number
$n=12$, $I=6$,    $D_I=37$.

 \vspace{0.1in}

\begin{center}
\includegraphics[width = 3in]{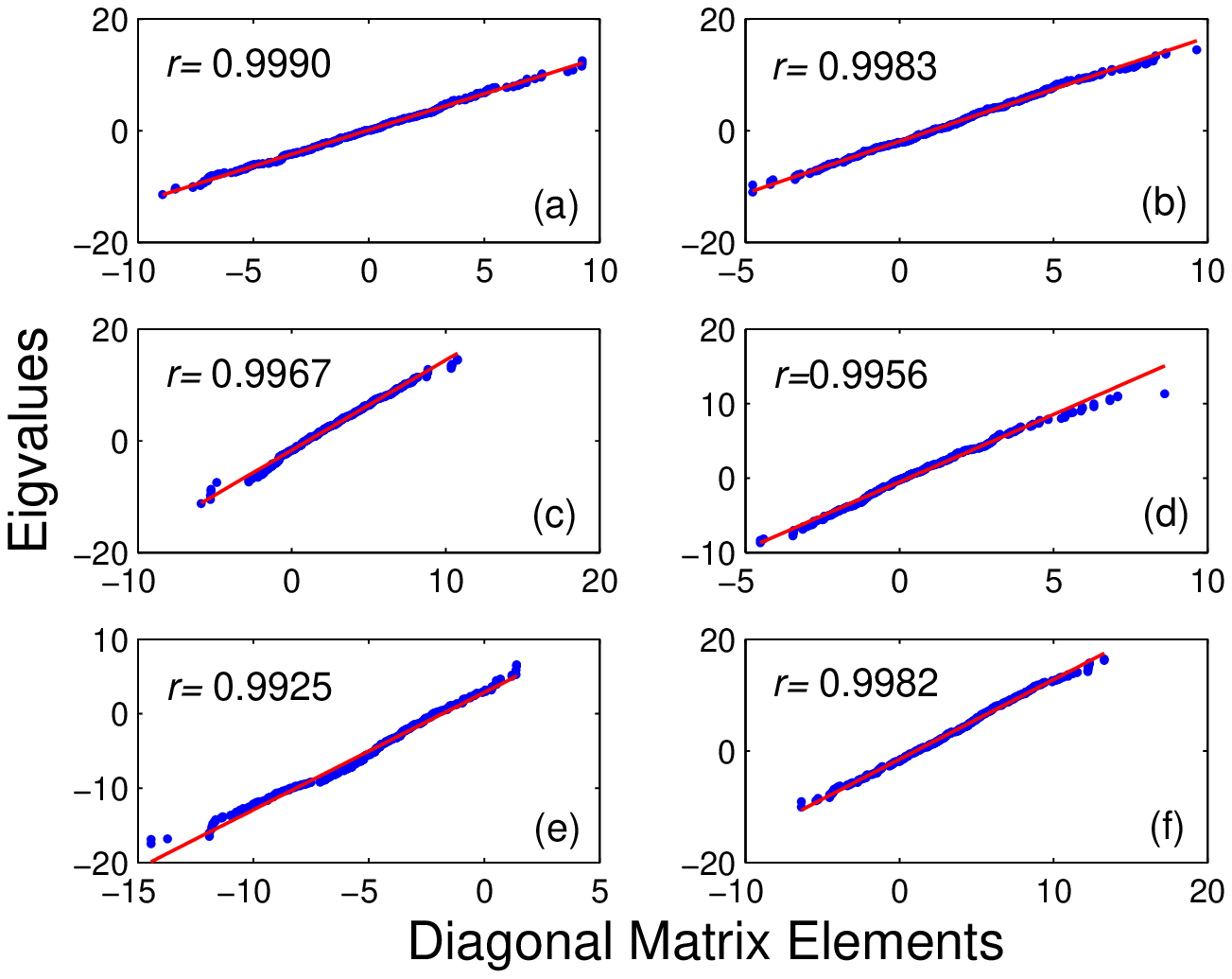}
\end{center}

FIG. 3. ~(Color online)~ The same as Fig. 1 except that these
examples correspond to three valence protons and three valence
neutrons in the $(2s_{1/2},1d_{3/2},0i_{11/2})$ shell. Here  $I=17$
and  $D_I=508$.

In the above examples the two-body interacting parameters are given by
Gaussian random numbers. The matrices of these systems are given by
coefficients of fractional parentage determined by geometry of the
configuration and two-body interactions which are taken to be
Gaussian. Now we come to matrices in which all matrix elements are
given by random numbers.

In the first set of examples, all matrix elements are taken to be
uniformly distributed random numbers between $-1$ and 1. Figure 4
shows  exact eigenvalues versus diagonal matrix elements (dimension
$D$=500), both sorted from smaller to larger values. The correlation
is seen to be very good, except for small deviations near the
largest and the smallest eigenvalues.

\vspace{0.1in}

\begin{center}
\includegraphics[width = 3in]{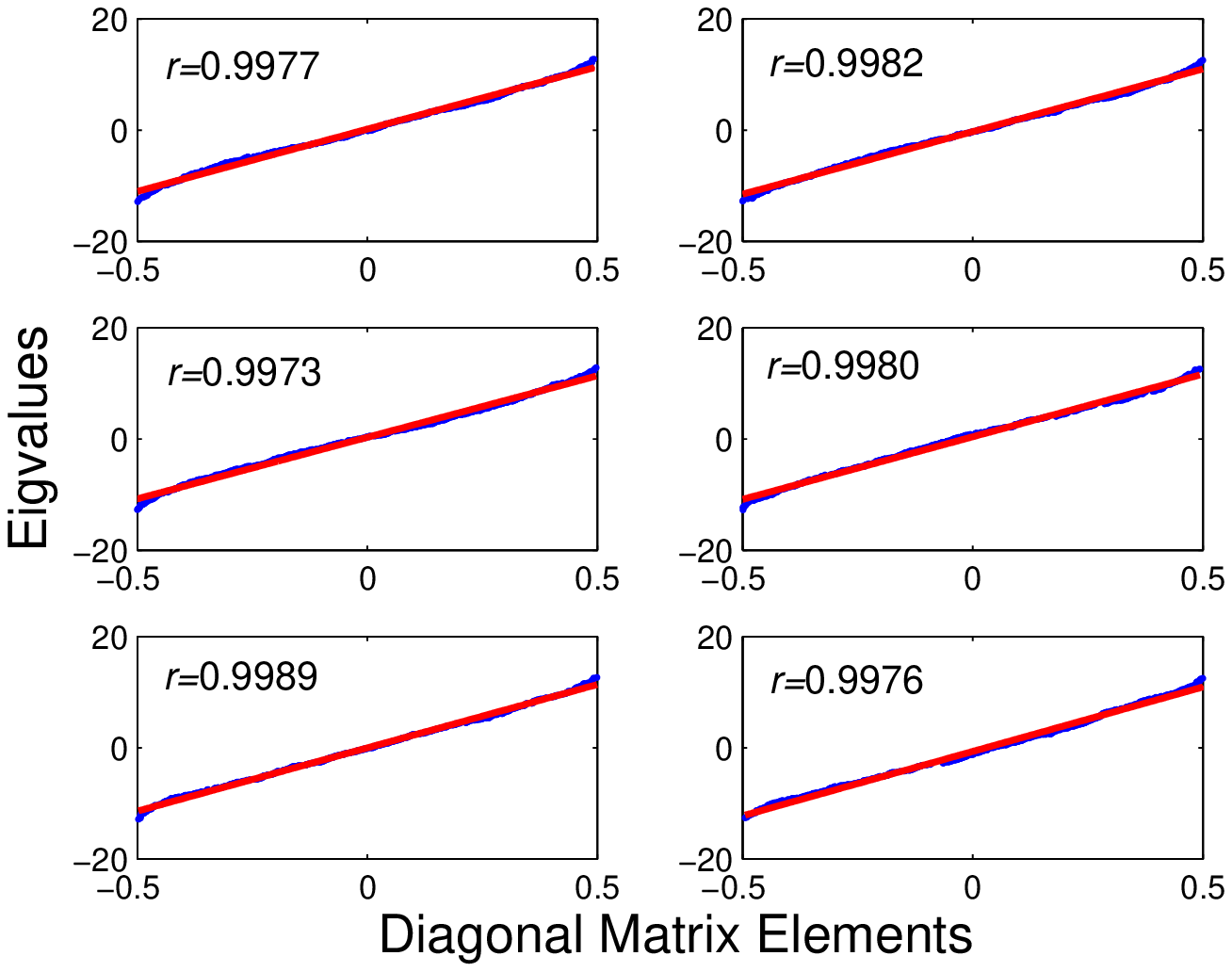}
\end{center}

FIG. 4  ~~ (color online)  ~ Eigenvalues versus  diagonal matrix
elements ($d_I=500$). Both eigenvalues and diagonal matrix elements
are sorted from smaller values to larger ones. All matrix elements
are taken to be uniformly distributed random numbers. One sees that
the eigenvalues predicted by linear correlation and the exact values
are almost equal except for a very few around the maximum and
minimum values.

\vspace{0.1in}

We also investigate the emergence of the above correlation in
matrices for which all matrix elements are random numbers following
the Gaussian  distribution.  Figure 5 shows results of matrices with
$D=500$ and all matrix elements being Gaussian random numbers. The
correlation seems not linear here, but rather a hyperbolic tangent
function because eigenvalues in these case follow  Wigner's
semicircle distribution. However, it is important to point out two
important facts. (I)~ There also exists a remarkable linear
correlation between eigenvalues and diagonal matrix elements on
average in these cases. The above linear correlation is actually
very good except for eigenvalues near the smallest or largest ones,
because the level density of eigenvalues in the middle part is much
larger than that near the minimum or maximum. {\it All} ``medium"
eigenvalues can be predicted very well by the linear correlation
determined by Eqs. (1-3). For matrices with dimension larger than
20, correlation coefficient $r$ is larger than 0.95. We note without
details that one easily obtains the hyperbolic tangent function
which seems proper in fitting the results around the minimum or
maximum eigenvalues, by using matrix elements in the same way as we
do for the TBRE systems. (II) ~ The correlation coefficients between
eigenvalues and diagonal matrix elements are larger than 0.995 if
one {\it artificially} enlarges diagonal matrix elements by about
two orders or more (typically by a factor of 50-60) for Gaussian
random matrices. The latter fact is very important, because in the
nuclear shell model calculations, magnitudes of diagonal matrix
elements are usually much larger than (non-zero) off-diagonal matrix
elements by about one or two orders.

\begin{center}
\includegraphics[width = 3in]{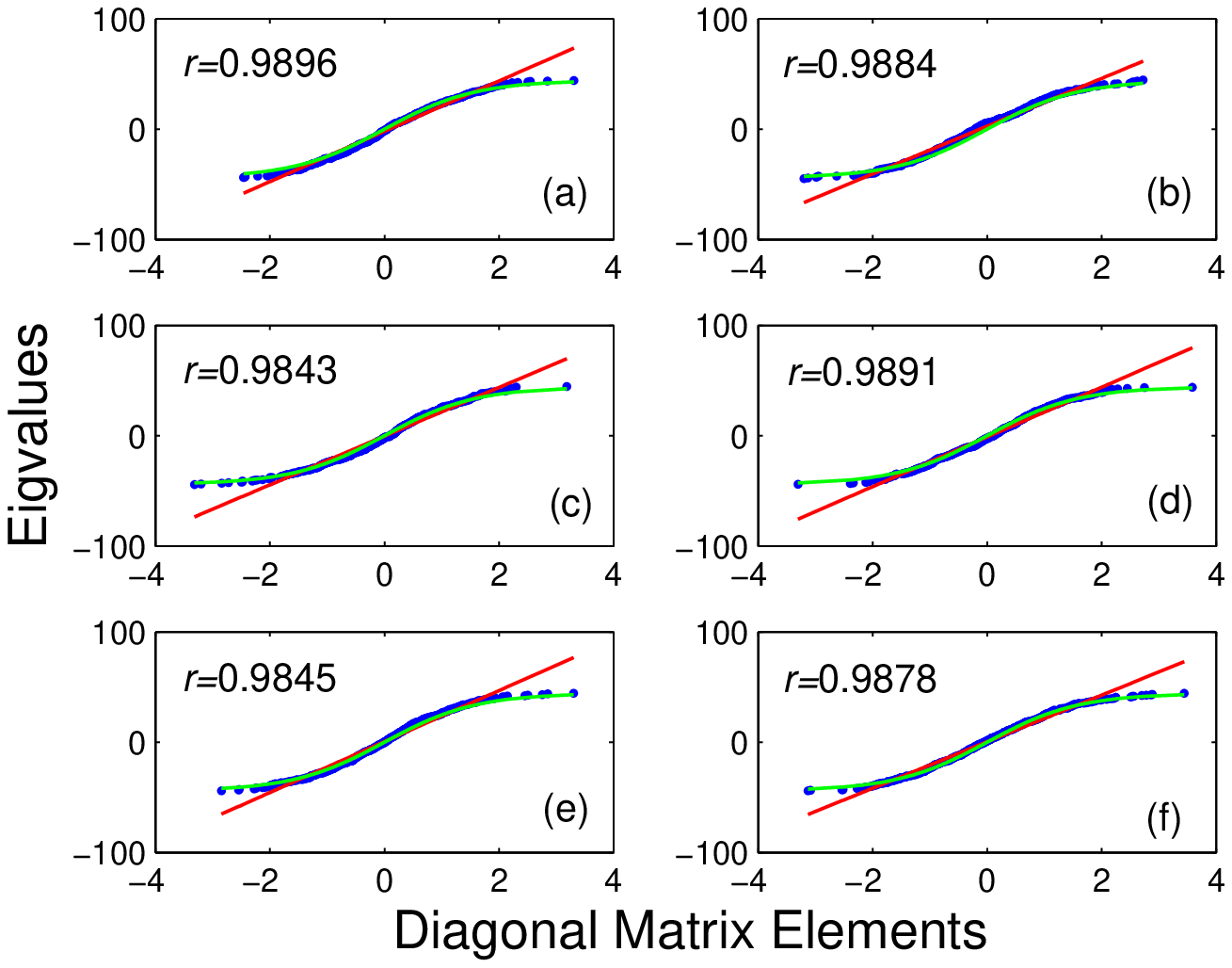}
\end{center}

FIG. 5. ~(Color online)~ Correlation between eigenvalues of random
matrices (Gaussian type) and diagonal matrix elements. The dimension
is 500 here. It is noted that the linear correlation is actually
very good except for eigenvalues near the smallest or largest ones,
because the level density of eigenvalues in the middle region is
much larger than that of eigenvalues near the minimum or maximum.
All $r$ values are larger than 0.98.

\vspace{0.2in}

Now we exemplify our approach by a realistic example. In Fig. 6(a),
we present eigenvalues of  spin $I=0$ states of $^{24}$Mg
($D_I=1161$) versus the diagonal matrix elements of the same
configurations. Here we use the USD interactions \cite{USD}. It is
seen that the linear correlation is very good. Similar results can
be found for other spin states. In Fig. 6(b) we show exact
eigenvalues $E_i^{\rm exact}$ in comparison with those predicted by
using linear correlations ($E_i$ in Eq.(1)). Although eigenvalues
predicted by using the  linear correlation deviates from the exact
values slightly, they are close to the exact ones in general. More
important, the agreement between predicted eigenvalues and exact
results is substantially improved when one goes to the medium
region. As shown in Fig. 6 (c), predicted 100th-105th $0^+$, $2^+$,
$3^+$, $4^+$, $6^+$ states are very close to exact eigenvalues
obtained by diagonalizations.

\vspace{0.2in}

\begin{center}
\includegraphics[width = 3.5in]{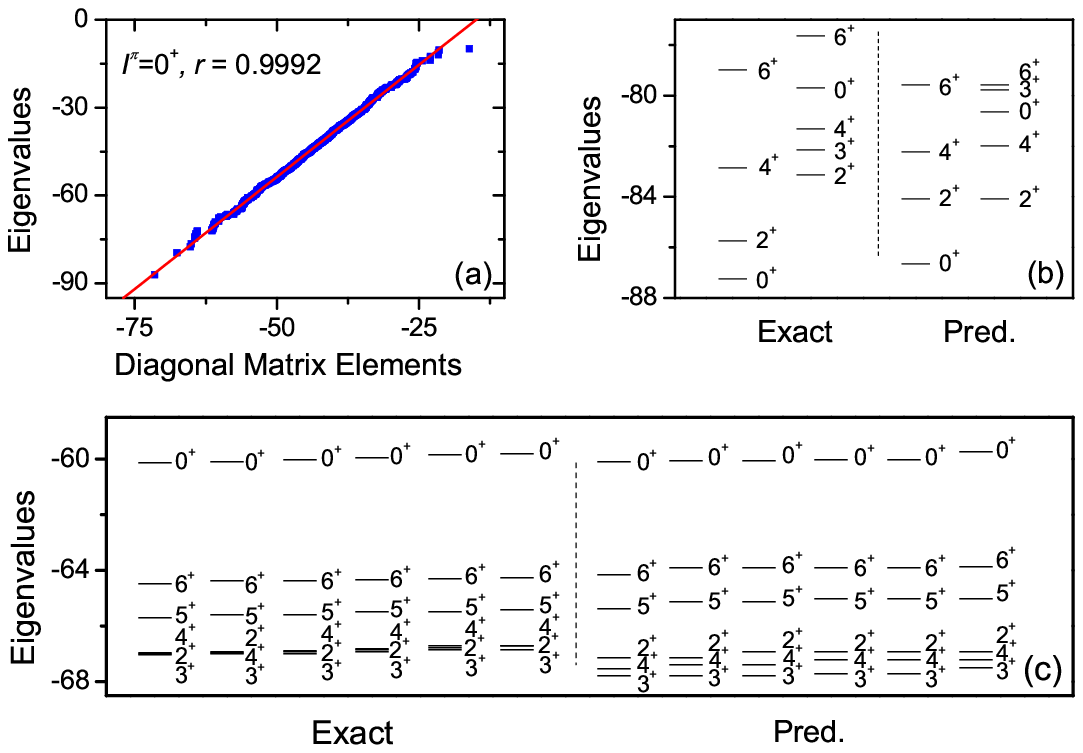}
\end{center}

\vspace{0.2in}

FIG. 6 ~~ Energy Spectrum for $^{24}$Mg. The USD interactions are
used here. (a) ~Eigenvalues of $I=0$ states obtained by
diagonalization versus diagonal matrix elements. One sees a striking
linear correlation. The correlation coefficient $r$ is larger than
0.999. (b) Energy spectrum by exact diagonalization and that
predicted by linear correlation. The eigenvalues of the lowest two
$I=2$ levels (and two lowest $I=6$ levels) are degenerate because
diagonal matrix elements corresponding to these states are equal due
to the isospin conservation of the USD interactions. (c) Energies of
excited states in medium region. The left hand side corresponds to
exact energies obtained by diagonalizations, and the right hand side
corresponds to predicted values by using linear correlations. The
100th-105th $0^+$, $2^+$, $3^+$, $4^+$, $6^+$ states are presented.

 \vspace{0.2in}

To summarize, in this paper we have discovered the linear
correlation between exact eigenvalues and diagonal matrix elements
for {\it individual} runs of the two-body random ensemble or  random
matrices, if both exact eigenvalues and diagonal matrix elements are
sorted from the smaller values to larger ones. We exemplify this
correlation by fermions in a single-$j$ shell, many-$j$ shells and
$sd$ bosons by using random interactions, and energy spectra of
$^{24}$Mg by using the USD interactions.

This linear correlation seems to be universal. For two-body random
ensembles, the linear correlation coefficient is $\sim$0.95-1.00;
for matrices in which all matrix elements are uniformly distributed
random numbers, the linear correlation is very good; for Gaussian
type random matrices,  there are sizable deviations between exact
eigenvalues and predicted eigenvalues  only if eigenvalues are close
to the minimum or maximum, and in this case the linear correlation
is well applicable to predicting levels in the medium region.

Although eigenvalues obtained here are not exact, they are very good
approximations of exact values. We therefore believe our results are
very useful in studying problems in which one needs approximate
eigenvalues when other methods are not possible.

 {\bf Acknowledgements:} The authors would like to
thank Prof. K. Ogawa for modifying his shell model code. Two of the
authors (JJS and YMZ) thank the National Natural Science Foundation
of China for supporting this work under grants 10575070, 10675081.
This work is also supported partly by the Research Foundation for
Doctoral Program of Higher Education in China under grant No.
20060248050, Scientific Research Foundation of Ministry of Education
in China for Returned Scholars, the NCET-07-0557, and  by Chinese
Major State Basic Research Developing Program under Grant
2007CB815000.

\end{document}